\documentclass[lettersize,journal]{IEEEtran}
\usepackage{amsmath,amsfonts}
\usepackage{amssymb}

\usepackage{algorithmic}
\usepackage{algorithm}
\usepackage{array}
\usepackage[caption=false,font=normalsize,labelfont=sf,textfont=sf]{subfig}
\usepackage{textcomp}
\usepackage{stfloats}
\usepackage{url}
\usepackage{verbatim}
\usepackage{graphicx}
\usepackage{cite}
\usepackage{url}
\usepackage{dsfont}

\usepackage{cuted}
\usepackage{multicol}

\usepackage[breaklinks]{hyperref}
\usepackage{enumitem}
\usepackage[export]{adjustbox}

\newcommand{\matr}[1]{{\boldsymbol{#1}}}
\renewcommand{\vec}[1]{{\boldsymbol{#1}}}
\newcommand{\EE}{\mathcal{E}}
\newcommand{\VV}{\mathcal{V}}
\newcommand{\GG}{\mathcal{G}}
\newcommand{\paulix}{\sigma^{(1)}}

\newcommand{\refequ}[1]{Eq.~(\ref{#1})}
\newcommand{\reffig}[1]{Fig.~\ref{#1}}

\usepackage[nolist]{acronym}
\begin{acronym}
\acro{FLOP}{floating point operation}
\acro{IVP}{initial value problem}
\acro{DAE}{differential algebraic system of equations}
\acroplural{DAEs}{differential algebraic systems of equations}
\acro{MNA}{modified nodal analysis}
\acro{QUBO}{quadratic unconstrained binary optimization}
\acro{QA}{quantum annealing}
\end{acronym}

\begin{document}

\title{Quantum Annealing based Power Grid Partitioning for Parallel Simulation}

\author{\IEEEauthorblockN{Carsten Hartmann\IEEEauthorrefmark{1}\href{https://orcid.org/0009-0007-5067-589X}{\includegraphics[width=3.2mm]{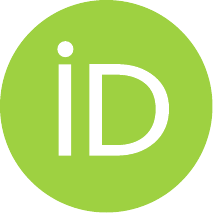}}, Junjie Zhang\IEEEauthorrefmark{1}\textsuperscript{,}\IEEEauthorrefmark{2}\href{https://orcid.org/0000-0003-1720-028X}{\includegraphics[width=3.2mm]{orcid.pdf}}, Carlos D. Gonzalez Calaza\IEEEauthorrefmark{4}\textsuperscript{,}\IEEEauthorrefmark{2}\href{https://orcid.org/0000-0001-8467-8633}{\includegraphics[width=3.2mm]{orcid.pdf}}, Thiemo Pesch\IEEEauthorrefmark{1}\href{ https://orcid.org/0000-0002-3297-6599}{\includegraphics[width=3.2mm]{orcid.pdf}} ,\\ Kristel Michielsen \IEEEauthorrefmark{4}\textsuperscript{,}\IEEEauthorrefmark{2}\href{https://orcid.org/0000-0003-1444-4262}{\includegraphics[width=3.2mm]{orcid.pdf}} and Andrea Benigni\IEEEauthorrefmark{1}\textsuperscript{,}\IEEEauthorrefmark{2}\textsuperscript{,}\IEEEauthorrefmark{3},\href{https://orcid.org/0000-0002-2475-7003}{\includegraphics[width=3.2mm]{orcid.pdf}}~\IEEEmembership{Senior Member,~IEEE}}

\IEEEauthorblockA{\IEEEauthorrefmark{1} ICE-1: Energy Systems Engineering\\ 
\IEEEauthorblockA{\IEEEauthorrefmark{4} Institute for Advanced Simulation, Jülich Supercomputing Centre }\\
Forschungszentrum Jülich, 52425 Jülich, Germany \\
\IEEEauthorblockA{\IEEEauthorrefmark{2} RWTH Aachen University, 52056 Aachen, Germany}\\
\IEEEauthorblockA{\IEEEauthorrefmark{3} JARA-Energy, Jülich 52425, Germany}\\
\{c.hartmann, a.benigni\}@fz-juelich.de}\\

}



\maketitle

\begin{abstract}

Graph partitioning has many applications in power systems from decentralized state estimation to parallel simulation. Focusing on parallel simulation, optimal grid partitioning minimizes the idle time caused by different simulation times for the sub-networks and their components and reduces the overhead required to simulate the cuts. Partitioning a graph into two parts such that, for example, the cut is minimal and the sub-graphs have equal size is an NP-hard problem. In this paper we show how optimal partitioning of a graph can be obtained using \ac{QA}. We show how to map the requirements for optimal splitting to a \ac{QUBO} formulation and test the proposed formulation using a current D-Wave QPU. 
We show that the necessity to find an embedding of the \ac{QUBO} on current D-Wave QPUs limits the problem size to under 200 buses and notably affects the time-to-solution. We finally discuss the implications of quantum hardware non-ideality on near term implementation in the simulation loop.

\end{abstract}


\section{Introduction}
The transition of energy systems towards renewable energies has significantly increased the complexity of modeling, simulating and controlling modern systems. Historically, power supply was adjusted to meet demand using a limited number of large power plants. However, in the future, power balancing is expected to be driven by a much larger number of smaller generation units, predominantly based on fluctuating renewable energy sources, alongside an increasing number of flexible consumers and prosumers in a grid with many more active grid elements. One approach to addressing the complexity of simulating and controlling these systems is to apply a parallelization strategy, which involves dividing the overall system into smaller subsystems that can be solved on separate computing nodes. This network partitioning can also facilitate the implementation of flexible, distributed and adaptive smart grid monitoring \cite{weiqing2008statediakoptics} and control solutions \cite{garcia2014clustering}. 

Other grid partitioning applications include dividing the power grid into optimally balanced regions to reduce i.\,e. administrative and computational complexity, and to simultaneously minimize loop flows \cite{sanchez2013partitioning}. Additionally, such partitioning can be applied to establish self-sufficient regions that allow for a certain level of electricity sharing \cite{tanjo2016graph}, enable controlled islanding to prevent the spread of failures \cite{sanchez_gracia2014spectralclustering}, and reduce the computational load in decentralized voltage control \cite{mori1993simulatedannealing} and diakoptics-based state estimation \cite{weiqing2008statediakoptics}. 

The focus of this paper is the problem of partitioning a power system graph to enable parallel simulation. While dependent on the specific simulation algorithm employed, in general \cite{armstrongMultilevelMATEEfficient2006, Dufour2011, Benigni2015}, an optimal partitioning for parallel simulation should ensure an adequate balancing of the computing loads and should minimize the overheads created by the partitioning. 

The problem of dividing a grid or graph into two or more parts is a fundamental problem in computer science and combinatorial optimization, known for its NP-hard complexity~\cite{kar_1972_NP_hard}. These problems are particularly difficult for classical computers to solve, which is why heuristics are usually applied~\cite{kyesswa_2020_partioning}. 
Consequently, recent advancements in quantum hardware have raised interest in exploring the potential of quantum computation as an alternative to traditional classical approaches. This has led to the emergence of a new research field: Quantum computation for power systems~\cite{zhou2022review, morstyn2024review}. 

Recent advances in this field can be grouped into three categories: 

First, classical algorithms to solve systems of linear or nonlinear equations are being replaced with quantum subroutines such as the HHL algorithm~\cite{harrow2009quantum} or specially designed variational quantum circuits~\cite{cerezo2021variational}. The HHL algorithm offers the potential for exponential speed-up in the asymptotic limit, provided the problem is sparse~\cite{harrow2009quantum} and the condition number can be tightly bounded~\cite{childs2017quantum}. However, it still faces challenges related to input/output processing and requires circuit depths that exceed the capabilities of current NISQ hardware. Despite these limitations, the HHL algorithm has been proposed for applications such as solving DC and AC power flow equations~\cite{eskandarpour2020quantum, feng2021quantum}. To circumvent some issues with the HHL algorithm, a variational circuit for a quantum electromagnetic transients program has been proposed \cite{zhou2022noisy}. However, quantum advantage has yet not been experimentally demonstrated for any variational approach \cite{cerezo2021variational}. 

Second, decision problems that represent fundamental operational or planning problems in power systems are being reformulated to fit quantum optimization approaches. These include the unit commitment (UC) problem \cite{ajagekar2019quantum_energy}, network reconfiguration in distribution grids \cite{silva2022qubo}, optimal power flow \cite{morstyn2022annealing}, and the aforementioned \cite{tanjo2016graph} grid partitioning into self-sufficient subgrids \cite{colucci2023partpowersystem}. However, on current hardware, only small-scale problems can be tackled, which spiked recent interest in the decomposition of the UC problem into smaller instances \cite{nikmehr2022unit_quantum, feng2022unit_quantum}.

Third, quantum optimization is being used to support classical simulation or optimization, for example, by finding optimal cuts in Benders decomposition to solve large instances of the UC problem~\cite{paterakis2023hybrid}.

The idea presented in this paper falls into the third category and explores graph partitioning to enhance the parallel simulation of power systems. The graph partitioning problem involves dividing the vertices of a graph into subgraphs while minimizing the number of edges between different subgraphs and/or achieving other specific objectives. Partitioning a graph into two parts, such that both subgraphs are of equal size, naturally lends itself to a \acf{QUBO} formulation~\cite{lucas_ising_2014}. Thus, the problem fits established quantum optimization paradigms such as \acf{QA} \cite{farhi_quantum_2000} or QAOA \cite{farhi2014quantum}. If hardware advances are realized, both approaches promise to solve large problem instances that are classically intractable. Partitioning into more than two parts, based on the concept of network modularity, can also be mapped to a \ac{QUBO} but it requires costly integer encoding~\cite{wangQuantumAnnealingInteger2023}. This work focuses on partitioning a graph into two parts. 

The article is structured as follows: 
Section ~\ref{sec:par:quantum_annealing_general} gives a short overview of \ac{QA}. 
Section~\ref{sec:par:quantum_partitioning_general} reviews, combines and extends QUBO formulations for graph partitioning into two parts. Section~\ref{sec:par:parallel_simulation_rev} gives a short overview of parallel simulation of power grids and its relation to graph partitioning, considering the specific algorithm of \cite{armstrongMultilevelMATEEfficient2006}. 
Section~\ref{sec:par:partitioning_for_HPC_formulation} combines the findings of the previous two sections to derive a QUBO for grid partitioning into two parts tailored to parallel simulation. We then briefly discuss possible extensions for partitioning into more than two parts. 
Section~\ref{sec:par:results} discusses the optimal and near-optimal solutions and current hardware limitations.

\section{Solving QUBOs using Quantum Annealing}
\label{sec:par:quantum_annealing_general}
\IEEEpubidadjcol

\subsection{Quantum Annealing in a Nutshell}

A \acf{QUBO} problem in its general form  is given by
\begin{align}
\begin{split}
    &\min_{\{z_n \in [0,1]\}} Q \\
    \label{eq:par:QUBO_general}
    &Q = \sum_{n,m = 1}^{M} z_n Q_{n,m} z_m,    
\end{split}
\end{align}
where $Q_{n,m}$ are the entries of an $M \times M$ symmetric matrix. 
The objective function of a \ac{QUBO} can be transformed into a classical Ising Hamiltonian $H_P$ by a linear transformation in the variables $s_n = 2z_n - 1 \in \{-1, 1\}$. $H_P$ is then the energy function of a spin glass, and the problem of finding an optimal solution of $Q$ has been translated into finding a ground state of $H_P$. Due to symmetries in the problem, the ground state energy can be degenerate and several ground states encode equivalent solutions.   

The idea of \ac{QA}\cite{farhi_quantum_2000} is based on the quantum adiabatic theorem \cite{born_beweis_1928, albash_adiabatic_2018} that roughly states that a closed system stays in its ground state if it undergoes an adiabatic transformation, that is a transformation that takes significantly longer than the time scales set by the internal dynamics. In quantum annealing, a system is initially prepared in a ground state of a mixer Hamiltonian $H_M$. Then the instantaneous Hamiltonian is ``slowly" evolved towards the problem Hamiltonian $H_P$ in an \emph{annealing schedule} given by 
\begin{align}
\begin{split}\label{eq:par:annealing_schedule}
    &H(t) = A(t) H_M + B(t) H_P, \\
    \text{with: } &A(0) \gg B(0), \quad A(T_A) \ll B(T_A),
\end{split}
\end{align}
where the tunneling energy $A$ and problem energy $B$ are monotonous functions whose typical schedule can be found in \cite{andriyash2016} and $T_A$ is the \emph{annealing time}, which should be large compared to the inverse of the minimum energy gap \cite{jansen_bounds_2007}, i.e. the energy difference between the ground state and the first excited state at the point of the evolution where these are closest in energy. 

For current D-Wave Advantage QPUs, the mixer $H_M$ is fixed as $\sum_n \paulix_n$ \cite{andriyash2016, lanting_experimental_2017}. Then, the ground state of $H_M$, thus the initial state, is given by the uniform superposition of all possible configurations in the eigenbasis of $H_P$. During anneal, $H_M$ causes quantum tunneling between configurations in analogy to thermal fluctuations in simulated annealing \cite{kadowaki_quantum_1998}. After the anneal, the state of the spin glass is measured in the eigenbasis of $H_P$.  
In an ideal \ac{QA} process, using a sufficiently large annealing time, the outcome $s_1...s_N$ of this measurement should be one of the ground states of $H_P$. 

Quantum annealing is a \emph{heuristic} solver as the adiabatic theorem does not strictly apply in practice. The real-world implementation is neither closed nor are the annealing times infinitely long. Consequently, the system can undergo a transition to an excited state of $H_P$ during anneal at any time, for example, due to thermal fluctuations. Then the outcome of an anneal is not a globally optimal solution to the original problem $Q$. 

For a more detailed introduction to \ac{QA} and a discussion of industry applications, we refer to \cite{yarkoni2022quantum}. 

\subsection{Assessing the Quality of a Solution}

Since \ac{QA} is a heuristic solver, similar to simulated annealing, one usually performs $N_s$ anneals Eq.~\eqref{eq:par:annealing_schedule} to create $N_s$ solution candidates. For each sampled configuration with energy $E_{sampled}$, the relative error is defined as 
\begin{equation}
    \label{eq:par:rel_error}
    L = \frac{\lvert E_{min, \, global} - E_{sampled} \rvert}{\lvert E_{min, \, global} \rvert}, 
\end{equation}
where $E_{min, \, global}$ is the energy of the global optimal solution.  In general, $E_{min, \, global}$ is not known a priori, except for two cases: The problem instance is small enough to be solved by brute force methods on CPUs, or the problem is formulated in such a way that the minimal energy of the objective function is known, cf., for example, the number partitioning problem discussed in \cite{lucas_ising_2014}.

To quantify the quality of a sample set, one defines the \emph{lowest relative error}\cite{mcgeoch_d-wave_2020} as
\begin{equation} 
    \label{eq:par:lowest_rel_error}
    L_{min} := \min_{E_{sampled}} L \,. 
\end{equation}

\subsection{Time to (Best) Solution}

As a consequence of sampling $N_s$ times, the annealing time $T_A$ is not the only parameter that affects the time to find a globally (or near) optimal solution, and there is a trade-off between longer annealing times and sampling more configurations. This trade-off is captured by the \emph{time to solution} ($TTS$) \cite{mcgeoch_d-wave_2020, mcgeoch2023milestonesquantumutilityhighway} which is defined as the ``time" to reach a globally optimal (or best) solution, at least once with probability $p_S$ and is given by
\begin{equation}
\label{eq:par:TTS}
    TTS(p_S, T_A) := \frac{\log (1-p_S)}{\log (1- \frac{N_{opt}}{N_s})} 
    T_A,
\end{equation}
where $N_{opt}$ is the number of globally optimal (or: best) solutions in the sample set. In practice one assumes that the ``success probability" of the heuristic solver, e.g. the quantum annealer, is below 1, that is, $N_{opt} < N_s$. Hence, Eq.~\ref{eq:par:TTS} is well defined. 

The time to solution only partially reflects the actual QPU access time~\cite{dwavetime2024}, as it only considers the annealing time and neglects delays and constant-overhead QPU programming and readout times. 

\subsection{Minor Embeddings}

The current D-Wave Advantage system has more than 5000 qubits and more than 35000 couplers \cite{mcgeoch_d-wave_2020} arranged in a Pegasus graph, in which qubits are coupled to at most 15 other qubits. To sample the \ac{QUBO} on the QPU, the \emph{logical graph} representing the couplings $Q_{n,m}$ between the variables $z_n$ and $z_m$ must be mapped to the hardware graph. If the \ac{QUBO} can not be mapped directly, for example, due to the \ac{QUBO} graph having a higher degree than that of the hardware graph, several physical qubits are strongly coupled ferromagnetically to form a \emph{chain} representing one logical variable (or qubit) \cite{choi_minor-embedding_2008}. Finding such a \emph{(minor) embedding} of the logical graph onto the QPU graph is usually done heuristically \cite{cai_practical_2014}.

\section{\ac{QUBO} Formulation for Graph Partitioning}
\label{sec:par:quantum_partitioning_general}
\IEEEpubidadjcol

A \emph{partition} of a graph $\GG = (\VV, \EE)$, where $\VV$ is a set of vertices/nodes and $\EE \subseteq \VV \times \VV$ a set of edges connecting the nodes, is an allocation of nodes to two (or more) sub-graphs $\GG_i = (\VV_i, \EE_i) \subset \GG $ such that each vertex $v \in \VV$ is in only one sub-graph. That is, the vertex sets $\VV_i$ are mutually disjoint and the union of all vertex sets is the original vertex set. The \emph{cut} set is comprised by all edges $e = (v_i, v_j)$ that connect nodes $v_i \in \VV_i$ and $v_j \in \VV_j$ in different parts, that is, $i \neq j$. In this work, we primarily focus on the partitioning into two parts as this admits a straightforward binary representation of the problem. 

Graph partitioning is a task frequently encountered in real-world applications, e.g. community detection \cite{Newman_2006_modularity}, studying immunization strategies \cite{chen_2008_immunization}, and decomposition of quadratic pseudo-Boolean functions \cite{billionnet_1989_decomp}. However, graph partitioning is known to be NP-hard \cite{kar_1972_NP_hard}. Dedicated heuristics exist to find the optimal partitioning based on the goals of the application \cite{Buluc2016}, in particular for power grids \cite{kyesswa_2020_partioning}. 

For the partitioning of a graph into two parts, another approach is to formulate the goals as objectives of a \ac{QUBO}.
The $N = \lvert \VV \rvert$ binary/decision variables $z_n$ encode to which part (sub-graph after partitioning) the $n$-th node belongs, that is 
\begin{equation}
    z_n = \begin{cases}
        1 \quad \text{if node $n \in \VV_1$,} \\
        0 \quad \text{if node $n \in \VV_2$.}
    \end{cases}
\end{equation} 
Following the work of \cite{lucas_ising_2014} and \cite{Newman_2006_modularity}, we briefly review and extend \ac{QUBO} formulations for four different objectives: Minimizing the cut \cite{lucas_ising_2014}, and balancing the sizes $N_i = \lvert \VV_i \rvert$ of the two sub-graphs $\VV_1$ and $\VV_2$ \cite{lucas_ising_2014}, balancing the sum of the edge weights within both parts and maximizing the network modularity \cite{Newman_2006_modularity}.

\subsection{\ac{QUBO} Formulations for the Single Objectives}

\subsubsection{Minimal (Weighted) Cut} 

One potential objective is to minimize the number of edges between both parts. A \ac{QUBO} formulation for this goal has been derived by Lucas \cite{lucas_ising_2014}. Here we extend Lucas's formulation to find the minimal \emph{weighted} cut, a problem that is also associated with finding the maximal flow \cite{Ford_Fulkerson_1956}. Let $w_{n,m}$ be the weight associated with the edge $e = (n,m)$. We seek to minimize the sum of the weights over all edges in the cut set
\begin{align}
    \begin{split}
    \label{eq:par:Q_cut_not_standard}
    Q_{cut} &= 
    \sum_{(n, m) \in \EE}  w_{n, m } (-2 z_n z_m + z_n + z_m ).
    \end{split}
\end{align}
Note that $(-2 z_n z_m + z_n + z_m ) = 1$ if nodes $n$ and $m$ are in different sub-graphs and evaluates to $0$ if they are in the same part.  

Using the adjacency matrix $\matr A $ of the graph $\GG$, we can rewrite Eq.~\eqref{eq:par:Q_cut_not_standard} as
\begin{align}
    \begin{split}
    \label{eq:par:Q_cut}
        Q_{cut} &= \sum_{n, m \in \VV} z_n (- A_{n,m} w_{n,m} ) z_m \\
        & \, + \sum_{n \in \VV} z_n \left( \sum_{m \in \VV} A_{n,m} w_{n,m} \right) z_n ,
    \end{split}
\end{align}
to resemble the standard form~\eqref{eq:par:QUBO_general} and read off the matrix elements $(Q_{cut})_{n,m}$. 

\subsubsection{Balancing the Sizes of the Sub-Graphs}
\label{sec:par:balance_size}

Another potential goal is to balance between the sizes, i.e., the number of nodes $N_i$, of both sub-graphs. To derive the objective function in \ac{QUBO} form given by Lucas \cite{lucas_ising_2014}, we start with the hard constraint that is satisfied if both sides have the same number of nodes
\begin{equation*}
     \sum_n z_n =  \sum_n (1 - z_n).
\end{equation*}
By squaring the difference between both sides, we turn the hard constraint into a quadratic objective function 
\begin{equation}
    \label{eq:par:Q_size}
    Q_{size} = \left( \sum_n z_n  + \sum_n (z_n - 1)  \right)^2,
\end{equation}
that penalizes a partitioning if $N_1 \neq N_2$. 
By multiplying out, $Q_{size}$ can be brought into the standard \ac{QUBO} form, cf. Eq.~\eqref{eq:par:QUBO_general}. 

\subsubsection{Balancing the Sum of the Edge Weights}

Instead of balancing the size of the sub-graphs, we could also seek to balance the sums of the edge weights $w_{n,m}$ (to the power $p$) between the sub-graphs. As a hard constraint, this can be written as 
\begin{equation*}
    \label{eq:par:Q_weights}
        \sum_{n\in \VV} \underbrace{\left( \sum_{m \in \VV} \frac{1}{2} A_{n,m} w_{n,m}^p \right)}_{:= \alpha(n, p)} z_n = \sum_{n\in \VV} \alpha(n, p) (1-z_n), 
\end{equation*}
where we have introduced the term $\alpha(n,p)$ to shorten the notation. To avoid counting the edges within a sub-graph twice, we have multiplied both sides by $\frac{1}{2}$ and moved this factor into the definition of $\alpha(n,p)$. Note that the weights of the edges in the cut set are attributed to both parts and thus cancel each other out. The power $p$ can be used to tune the objective. If $p=1$, the sums of all weights in each part are balanced. For $p>>1$, the contributions of small weights are scaled down relatively to the maximal weights, such that balancing the maximal weights is emphasized. 

Again, this constraint can be turned into a quadratic objective function $Q_{weights}(p)$ by the same steps as for $Q_{size}$. We get 
\begin{equation}
    \label{eq:par:Q_weights}
    Q_{weights}(p) = \left(2 \sum_{n \in \VV} \alpha(n,p)z_n - \sum_{n \in \VV} \alpha(n,p) \right)^2.
\end{equation}

\subsubsection{Network Modularity}

Lastly, maximizing the network modularity \cite{Newman_2006_modularity} is a common method to detect communities, that is, highly connected components representing functional units. These components provide a natural partitioning of the graph. The (negative) network modularity is equivalent to an Ising Hamiltonian and, thus, admits a straightforward \ac{QUBO} formulation given by 
\begin{equation}
    \label{eq:par:Q_mod}
    Q_{mod}= \sum_{n,m} z_n \left(\frac{-1}{M} (A_{n,m} - \frac{deg(n) deg(m)}{2M}) \right) z_m, 
\end{equation}
where $M$ is the number of edges, $deg(n)$ the degree of node $n$. 
To derive $Q_{mod}$ from \cite{Newman_2006_modularity} we have used that $\sum_m A_{n,m} = deg(n)$ and $\sum_n deg(n) = 2M$ in order to eliminate the first order terms. 

Graph partitioning into more than two parts using integer encoding based on the idea of network modularity has been recently discussed by Wang et al. ~\cite{wangQuantumAnnealingInteger2023}.

\subsection{QUBO using Linear Scalarization}

Combining several of the previously introduced objectives, we obtain a \emph{multi-objective optimization} program with potentially \emph{conflicting objectives}. For example, $Q_{cut}$ is minimized if the cut set is empty; that is, all nodes are allocated to the same part. However, this is not optimal if we also want to, e.g., balance the sizes of the two sub-graphs. 

For quantum optimization, we need to turn the multi-objective optimization program into a single \ac{QUBO}. Hence, we collect the single objectives in a weighted sum 
\begin{align}
\begin{split}
    \label{eq:par:single_objective}
    Q = \sum_i \lambda_i Q_i,
\end{split}
\end{align}
where $i \in \{ cut, \, size, \, weights, \, mod \}$. 

The coefficients $\lambda_i$ have to be chosen a priori and effectively determine the importance of each objective as they set the energy scale. Different choices for the coefficients $\lambda_i$ can lead to different optimal points on the convex hull of the Pareto front \cite{das_closer_1997}. To solve the multi-objective optimization problem, one samples different Pareto optimal points using different sets of coefficients $\lambda_i$. The final solution is then selected by a ``decision maker" using domain knowledge of the problem \cite{miettinen1999nonlinear}.

\section{Parallel Simulation of Power Grids Based on network partitioning}
\label{sec:par:parallel_simulation_rev}
\IEEEpubidadjcol

Parallel simulation of power networks following partitioning has been a common technique in power system simulations for decades. There are different but similar approaches proposed for parallel simulation with partitioned networks, which all involve simulating the partitioned sub-networks in parallel with a sequential part left associated with the link between the sub-networks~\cite{happDiakopticsSolutionSystem1974, armstrongMultilevelMATEEfficient2006, shuParallelTransientStability2005}. 
Take the diakoptics-based method MATE~\cite{armstrongMultilevelMATEEfficient2006} as an example. The \ac{MNA} formulation of the electric network $YV=I$ for a partitioned network assumes a block diagonal form
\begin{equation}\label{eq:part_mate_1}
    \begin{bmatrix}
        \begin{bmatrix}
            Y_{N1} & 0 \\
            0 & Y_{N2}
        \end{bmatrix}
         & C \\
        C^{\top} & -Z_{L}
    \end{bmatrix}
    \begin{bmatrix}
        v_{N1} \\
        v_{N2} \\
        i_L
    \end{bmatrix}
    =
    \begin{bmatrix}
        h_{N1} \\
        h_{N2} \\
        -V_L
    \end{bmatrix},
\end{equation}
where $h_{Nk}$, $Y_{Nk}$, $v_{Nk}$ for $k\in \{1,2\}$ represent current injections, admittance matrix, nodal voltages (including currents chosen as independent variables as of \ac{MNA}) in each sub-networks, respectively. 
$C$ represents the connectivity matrix of the link between partitions (referred to as``cut" in the following), and $Z_L$, $i_L$, $V_L$ are the impedance matrix, current and voltage vectors over the cut, respectively. 

Consider the simplified system of equations
\begin{equation*}\label{eq:part_mate_2}
    \begin{bmatrix}
        Y & C \\
        C^{\top} & -Z_{L}
    \end{bmatrix}
    \begin{bmatrix}
        V \\
        I_L
    \end{bmatrix}
    =
    \begin{bmatrix}
        H \\
        -V_L
    \end{bmatrix}.
    \quad 
    \begin{matrix}
        \text{\textcircled{1}} \\ \text{\textcircled{2}}
    \end{matrix}
\end{equation*}
Adding $-C^{\top}Y^{-1} \cdot\text{\textcircled{1}}$ to \textcircled{2} and multiplying $\text{\textcircled{1}}$ with $Y^{-1}$ from the left yields
\begin{equation*}\label{eq:part_mate_3}
    \begin{bmatrix}
        1 & Y^{-1}C \\
        0 & -Z_{L} - C^{\top}Y^{-1} C
    \end{bmatrix}
    \begin{bmatrix}
        V \\
        I_L
    \end{bmatrix}
    =
    \begin{bmatrix}
        Y^{-1} H \\
        -V_L - C^{\top}Y^{-1} H
    \end{bmatrix}.
\end{equation*}
Hence, \refequ{eq:part_mate_1} can be reformulated as
\begin{equation}\label{eq:part_mate_4}
\begin{split}
    \begin{bmatrix}
        1 & 0 & Y_{N1}^{-1}C_{N1} \\
        0 & 1 & Y_{N2}^{-1}C_{N2} \\
        0 & 0 & C^{\top}_{N1}Y_{N1}^{-1}C_{N1} + C^{\top}_{N2}Y_{N2}^{-1}C_{N2} + Z_L \\
    \end{bmatrix}
    \begin{bmatrix}
        v_{N1} \\
        v_{N2} \\
        i_L
    \end{bmatrix}
    &= \\
    \begin{bmatrix}
        Y_{N1}^{-1}h_{N1} \\
        Y_{N2}^{-1}h_{N2} \\
          C_{N1}^{\top}Y^{-1}_{N1}h_{N1}+C_{N2}^{\top}Y^{-1}_{N2}h_{N2} + V_L
    \end{bmatrix} \, &
\end{split}
\end{equation}
which shows that the solutions of each sub-network become independent of each other except for the cut. Hence, parallel computation of the sub-networks is possible after solving the equations over the cut. Additionally, finer parallelizations can be applied to each sub-network, such as the approaches in~\cite{Dufour2011, Benigni2015, zhangShiftedFrequencyAnalysis2024, Aristidou2014}. Overall, each time step in simulating a partitioned network can be divided into three stages~\cite{armstrongMultilevelMATEEfficient2006}:
\begin{enumerate}
    \item Components of each sub-network are solved to obtain the injections to the network,
    \item Equations over the cut are solved,
    \item The network equations in each sub-network are solved in parallel with the injections from components and the cut.
\end{enumerate}

Based on these three steps, we can formulate an optimization problem to find good partitioning so that the overall computation time is minimized. The objectives for optimal partitioning are:
\begin{enumerate}
    \item minimizing the additional computational cost due to the cut, 
    \item minimizing the difference in computational cost among sub-networks to minimize the ``idle time".
\end{enumerate}

\section{Grid Partitioning to minimize overhead in parallel simulation}
\label{sec:par:partitioning_for_HPC_formulation}
\IEEEpubidadjcol

\subsection{Estimation and Balancing of Computational Costs}
\label{sec:par:comp_costs_steps}

We now need to estimate the computational costs of simulating the partitioned sub-networks to derive the objective function. 
For simplicity, we approximate these costs using the number of computations (in the form of \acp{FLOP}) needed to evaluate the equations. 

\subsubsection{Component costs}
The cost $c_i$ for solving the equations governing the $i$\textsuperscript{th} component, e.g. a generator, depends on the numerical integration method used to evaluate the derivative or Jacobian of the model. 
Let $\sigma_{i, j}$ be the number of \acp{FLOP} for evaluating the equation governing the $j$\textsuperscript{th} state variable in the $i$\textsuperscript{th} component model. 
We have
\begin{equation*}
    c_{i} = k\sum_j \sigma_{i,j} + \Delta_i, 
\end{equation*}
where $k$ represents the number of times the derivatives or the Jacobian is evaluated. That is, $k$ depends on the number of stages or iterations of the integration method. 
Here, $\Delta_i$ quantifies the additional overhead such as vector addition, e.g. in $x \leftarrow x+dx$. For simplicity, we imply the use of Euler forward and ignore the additional overhead in this work, which yields
\begin{equation*}
    c_{i} \approx \sum_j\sigma_{i,j}.
\end{equation*}
We can illustrate the derivation of $c$ via a simple classical generator model~\cite{kundurPowerSystemStability1994} with two state variables
\begin{align*}
    \dot{\delta} &= \omega_{b} (\omega - \omega_s), \\ 
    \dot{\omega} & = \frac{1}{2H}\left[(T_m-(\psi_d i_q - \psi_q i_d) \right].
\end{align*}
The cost $\sigma_{\dot{\delta}}$ to evaluate the first state variable $\dot{\delta}$ can then be derived by counting the \acp{FLOP}, which here is $2$ for one subtraction and one multiplication. 
Similarly, $\sigma_{\dot{\omega}}$ is 5, since $\frac{1}{2H}$ is considered as one constant. Thus, the cost $c$ is $c = \sigma_{\dot{\delta}} + \sigma_{\dot{\omega}} = 7$. 
The generator model implied in this work is a ninth-order model~\cite{kundurPowerSystemStability1994}. This model is significantly more complex, heuristically we evaluate the cost as $c= 350$.

A partitioning of the network into two sub-networks decomposes all components into three disjoint sets: the sets of all components in distinct parts (sub-network 1, 2), and in the cut, respectively. Let $C_{1(2)}$ and $C_{cut}$ be the index sets, e.g. the i\textsuperscript{th}  component is in part 1 if $i \in C_1$. 
Then the ``idle time", i.\,e. mismatch of computing load in different sub-networks, 
is given by 
\begin{equation}
     \label{eq:par:component_idle_time}
    \Delta W_{comp.} = \sum_{i \in C_1} c_i - \sum_{i \in C_2} c_i. 
\end{equation}

\subsubsection{Cut costs}

Computations on the cut consist of two parts, updating the mutual injections between the cut and sub-networks and solving the component equations and network equations within the cuts. Since the computation for updating the injections are matrix-vector multiplications, the cost for solving the component equations and updating the injections for two sub-networks is given by 
\begin{equation*}    
    W_{cut, \, inj.} = \sum_{i \in C_{cut}} c_i + 4 M_c (N-1) + M_c,
\end{equation*}
where $M_c$ is the number of edges in the cut and $N$ is the total number of nodes/busses. 

Solving for the cut or the sub-networks (see below) is equivalent to solving independent sets of linear equations, i.\,e. solving $Ax=b$ for $x$. We assume that LU decomposition is used since the decomposition must only be performed once if there are no topological changes. In each simulation step, the cost for solving the (sub)-network is then given by the \acp{FLOP} for forward and backward substitution. Hence, for solving the network equations over the cut, we have
\begin{equation*}    
    W_{cut, \, net.} = 2 M_c^2-M_c.
\end{equation*}
Using that $M_c \ll N$, the total overhead induced by the cut can be estimated as 
\begin{align}    
    W_{cut} \approx \sum_{i \in C_{cut}} c_i + 4  (N-1) M_c.
    \label{eq:par:cut_serial_time}
\end{align}

\subsubsection{Sub-network costs}

Solving for the sub-networks is again done by LU decomposition. Therefore, the mismatch in solving the sub-networks is given by the difference between forward and backward substitution costs in each part
\begin{align}
\begin{split}    
    \Delta W_{net.} &= 2(N_1^2 - N_2^2)-(N_1-N_2) \\
     &= 2N\Delta N-\Delta N .\label{eq:par:network_idle_time}
\end{split}    
\end{align}
where $N_{1 (2)}$ is the number of nodes/busses in sub-network $1$, or $2$ respectively.  

\subsection{Graph representation}\label{subsec:net_as_graph}

\begin{figure}
    \centering
    \includegraphics[width=.8\linewidth]{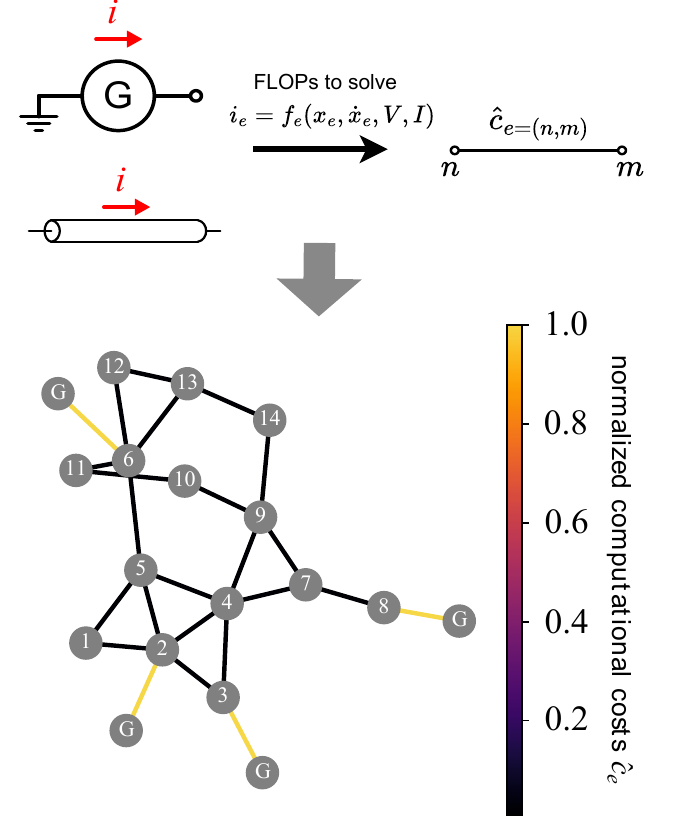}
    \caption{Steps to turn power grid into $\GG_{comp}$. First, map each component, e.g. generator and lines, to an edge $e$ whose weight $\hat{c}_e$ represents the \acp{FLOP} needed to solve for the current $i_e$. Second, build a graph that resembles the grid topology.  }
    \label{fig:par:problem_graph_case_14}
\end{figure}

Let $\GG_{grid} = (\VV_{grid}, \EE_{grid})$ be a graph representing the grid topology, that is, the nodes $n \in \VV_{grid} $ represent buses and the edges $e \in \EE_{grid}$ represent the electrical transmission lines. To put all the components, i. \,e. generators and transmission lines, on an equal footing, we construct a \emph{component graph} $\GG_{comp} = (\VV_{comp}, \EE_{comp}) \supset \GG_{grid}$ based on the grid $\GG_{grid}$ in the following way: Except the transmission lines that are already represented as edges, we represent each component attached to a bus $n \in \VV_{grid}$ by an \emph{additionally added} edge\footnote{We also add an additional node for each component.} that is incident to the bus $n$. Hence all electrical components are represented by edges in the graph $\GG_{comp}$. To all edges, we then assign a weight $\hat{c}_e = c_e/c_{max} \in [0,1]$ that represents the normalized computational cost (normalized \acp{FLOP}) of simulating the respective component, cf.~Section~\ref{sec:par:comp_costs_steps}. The steps to construct $\GG_{comp}$ are illustrated in Fig.~\ref{fig:par:problem_graph_case_14}. 

The problem of finding an optimal partitioning of the network into two sub-networks for parallel simulation is thus equivalent to finding an optimal partitioning of the graph $\GG_{grid}$ taking into account all components whose computational costs are represented by the weights $\hat{c}_e$ of the edges in $\GG_{comp}$. We only need to partition the graph $\GG_{grid} \subset \GG_{comp}$ as the nodes in $\VV_{comp} \setminus \VV_{grid}$ do not represent any additional partition choices regarding the network and thus the components: All edges representing components between nodes in different parts $\VV_i$ are in the cut set. For all other components, there is at least one node $n \in \VV_{grid}$ incident whose assignment to $\VV_1$ or $\VV_2$ provides the assignment of the component to part 1 or part 2.

\subsection{\ac{QUBO} Formulation}

To find an optimal partitioning of $\GG_{grid}$ we map each of the three objectives of minimizing the costs for simulating the cut, balancing the cost of simulating the components and the network in each part, see section~\ref{sec:par:parallel_simulation_rev}, to one of the general \ac{QUBO} objectives from section \ref{sec:par:quantum_partitioning_general}. Each of these mappings is chosen such that the objective values of the \ac{QUBO} term are some measure for the computational overhead that arises due to the partitioning into sub-networks, cf. section \ref{sec:par:comp_costs_steps}. The sum of all three objective values is then an estimate for the quality of the partitioning and the full \ac{QUBO} objective function is thus a simple sum of three terms
\begin{equation}
\label{eq:par:Qubo_grid_part}
    Q_{part.} =  Q_{comp.} + Q_{cut} + Q_{net.}.
\end{equation}
Due to this construction, we avoid the additional problem of choosing the weights $\lambda_i $ a priori. However, rescaling the objectives, like any reformulation, can affect the solution time if heuristics are involved without changing the solution qualitatively, see discussion in section~\ref{sec:par:rescaling_qubo}.

\subsubsection{Reduce ``idle" time in component simulation} 

For the component simulation step, we want to minimize the idle time that arises if one part can be simulated faster than the other, see Eq.~\eqref{eq:par:component_idle_time}. By adopting from $Q_{weights}(p=1)$ in Eq.~\eqref{eq:par:Q_weights}, a \ac{QUBO} objective function $Q_{comp.}$ that computes the squared idle time for a given partitioning is given by 
\begin{align}
\label{eq:par:QUBO_sum_costs}
        Q_{comp.} &= \left( 2 \sum_{n \in \VV_{grid}} z_n  \Tilde{\alpha}(n) - \sum_{n \in \VV_{grid}} \Tilde{\alpha}(n) \right)^2
\end{align}
with 
\begin{align*}
    \Tilde{\alpha}(n) = \frac{1}{2} \sum_{m \in \VV_{grid}} \hat{c}_{n,m} + \sum_{m \in \VV_{comp.} \setminus \VV_{grid.}}  \hat{c}_{n,m}.
\end{align*}
The definition of $\Tilde{\alpha}(n)$ takes into account edges in $\EE_{comp.} \setminus \EE_{grid.}$ are incident to only one node in $\VV_{grid}$ by construction, and thus, are not counted twice. On the other hand, all edges in $\EE_{grid.}$  are incident to two nodes in $\VV_{grid}$, and thus, a factor of $\frac{1}{2}$ is introduced in the first term of $\Tilde{\alpha}(n)$ to avoid double-counting. 

\subsubsection{Minimize the computational costs of cut simulation} 
\label{sec:par:minimize_cut_net_comp}

Minimizing the total overhead due to the cut simulation can be mapped to finding a weighted minimal cut. We have already introduced a \ac{QUBO} $Q_{cut}$ formulation for weighted minimal cuts in Eq.~\eqref{eq:par:Q_cut}. To incorporate the costs for the component and network simulation of the cut and compute the overhead according to Eq.~\eqref{eq:par:cut_serial_time}, we define the edge weights $w_{n,m}$ as 
\begin{equation}
    w_{n,m} = \hat{c}_{n,m} + 4 \frac{N-1}{c_{max}},
\end{equation}
for the edges in $\GG_{grid}$. Again, $N$ is the number of buses in the grid, thus $N = \lvert \VV_{grid} \rvert$. We normalize the costs attributed to the network and the injections by $c_{max}$ to obtain normalized flops. Then, the objective function that estimates the overhead due to the sequential simulation of the cut is given by 
\begin{align}
\begin{split}
        Q_{cut} = \frac{1}{2} \sum_{n, m \in \VV_{grid}}  A_{n,m}(\GG_{grid}) \left( \hat{c}_{n,m} + 4 \frac{N-1}{c_{max}}\right) \\ (-2 z_n z_m + z_n + z_m).  
\end{split}     
\end{align}

\subsubsection{Reduce ``idle" time in network simulation}

For the parallel network simulation step of the sub-networks, both parts should have the same size (=number of nodes) to reduce the idle time given by Eq.~\eqref{eq:par:network_idle_time}. We thus scale  $Q_{size}$ in Eq.~\eqref{eq:par:Q_size} as 
\begin{equation}
    \label{eq:par:QUBO_network_costs}
    Q_{net.} = \frac{(2N-1)^2}{c_{max}^2} \left( 2 \sum_{n \in \VV_{grid}} z_n - N \right) ^2
\end{equation}
in order to calculate the squared idle time. 

\subsection{Partitioning for more than two computational nodes}

Suppose we seek to partition the grid into $P$ parts. In that case, a direct approach is to define integer variables $z_n \in \{0, 1, \ldots P-1\}$ or one-hot encoded variables $z_{n,p}$, where $z_{n,p} = 1$ if and only if bus $n$ is in part $p$, to encode the partitioning and then extend the \ac{QUBO} objectives to the more general case. One-hot encoding has been used in the aforementioned \cite{colucci2023partpowersystem} to minimize the cuts and balance the sizes of the subgrids. However, both encodings significantly increase the complexity, in particular, the number of qubits. Hence, even small problem instances become impossible to embed and solve on near-term hardware. 

If we seek to distribute the computation to $P = 2^n$ workers, another approach is to partition the (sub-)networks into two parts iteratively. However, an iterative approach does not guarantee a globally optimal result. The difference between an iterative quantum optimization approach and the mixed integer programming approach introduced above is a research question that we intend to investigate in the near future. 

\subsection{Ensuring Connectedness}
\label{sec:par:ens-connectedness}

The derived \ac{QUBO} objective function $Q_{part.}$ does not enforce that each sub-network is connected. In case a sub-network is not connected, the costs for the simulation of the sub-network are overestimated in \refequ{eq:par:network_idle_time} because each connected part can be treated separately since the matrix $Y_{Ni}$ is itself also a block matrix. Nevertheless, disconnected solutions are feasible solutions for distributed parallel simulation.

Ensuring connectedness is a similar problem to the traveling salesperson problem (TSP), where a satisfactory solution describes a \emph{single} tour passing through all the cities that need to be visited. A known issue in the Dantzig-Fulkerson-Johnson formulation \cite{dantzig1954tsp} of the TSP is that solutions often contain sub-tours, i.e., tours through subsets of cities. An approach for dealing with these disconnected sub-tours is the ``finger-in-the-dike" method \cite{Monta_ez_Barrera_2023}, where the problem is initially solved without banning sub-tours, and ad-hoc constraints are introduced iteratively to the original formulation to ban them as they appear in the sampled solutions. Analogously, here, one could ban disconnected sub-networks as they are sampled to ensure the connectedness of the sub-networks.

\section{Results}
\label{sec:par:results}
\IEEEpubidadjcol

For reproducibility and the fact that most readers will be familiar, we use the IEEE14, 30, 57, 89, 118, 145, 200, 300~\cite{ExampleMatpowerCases} test grids to investigate optimal solutions, resource demands and solution quality on the QPU for $Q_{part.}$ 

\subsection{Optimal Solutions}

\begin{figure}[t]
    \centering
    \includegraphics[width=\columnwidth]{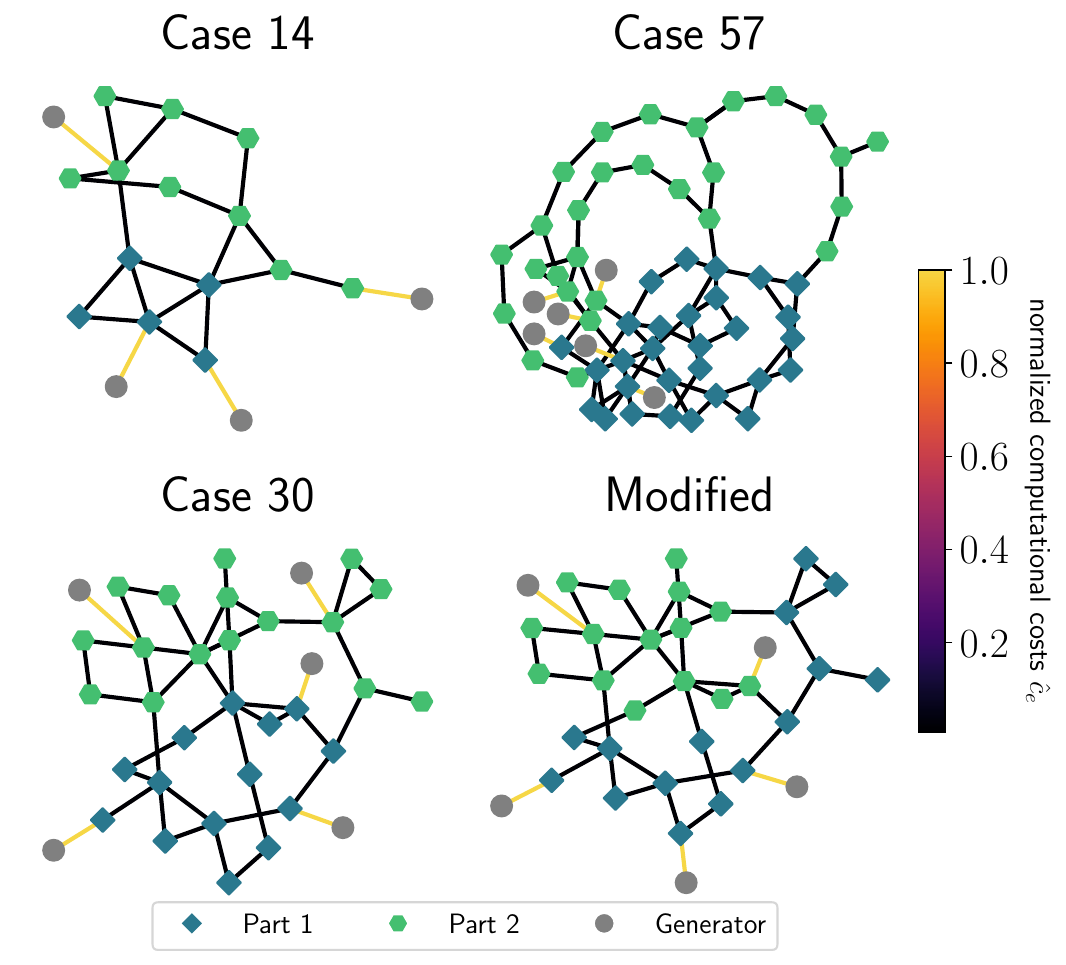}
    \caption{Optimal partitionings for Case 14, case 57, case 30 and modified case 30, where one generator has been moved to a different bus. The optimal solutions have been obtained using Gurobi. }
    \label{fig:par:case30_optimal_solutions}
\end{figure}

We use Gurobi on a Lenovo ThinkPad T14 with an Intel(R) Core(TM) i5-10210U CPU (4 cores, 8 Threads) and 32 GB installed RAM to find the reference solutions of $Q_{part.}$ for the test cases with 14, 30, and 57 buses. Gurobi uses a branch and bound method to solve mixed integer problems (MIP). It is a deterministic solver that uses heuristics, e.g. for node selection and cut selection. The optimal solutions (MIPGap = 0) are found in $T_{Gur.} = 0.43s$ for case 14, in $ T_{Gur.} =4.66s$ for case 30 and in $T_{Gur.} = 1.04s$ for case 57, see also Table \ref{tab:par:solver_comp} and the discussion in section \ref{sec:par:hyper} for a base comparison to the performance of \ac{QA}. 

The optimal solutions for the small test cases 14, 30 and 57 demonstrate that minimizing $Q_{part.}$ provides a reasonable partition into two sub-networks, cf. Fig.~\ref{fig:par:case30_optimal_solutions}. The sub-network sizes and the components are balanced and the cut sets are small. Furthermore, relocating a generator, though not a realistic scenario, demonstrates how modifications or reconfigurations of the network can result in different optimal partitionings for each configuration, see the lower right panel in Fig.~\ref{fig:par:case30_optimal_solutions}.  However, it should be noted that the partitioning itself introduces an overhead, necessitating the minimization of $Q_{part.}$ and the conduction of LU decomposition. Consequently, if a change is only persistent for a brief duration, repartitioning may result in an increase of the overall computational cost of the simulation. In our future work, we plan to investigate the possibility of including the overhead of repartitioning within the presented framework. 

\subsection{Scaling of Computational Resources}
\label{sec:par:scaling}

\begin{figure}[t]
    \centering
    \includegraphics[width=\columnwidth]{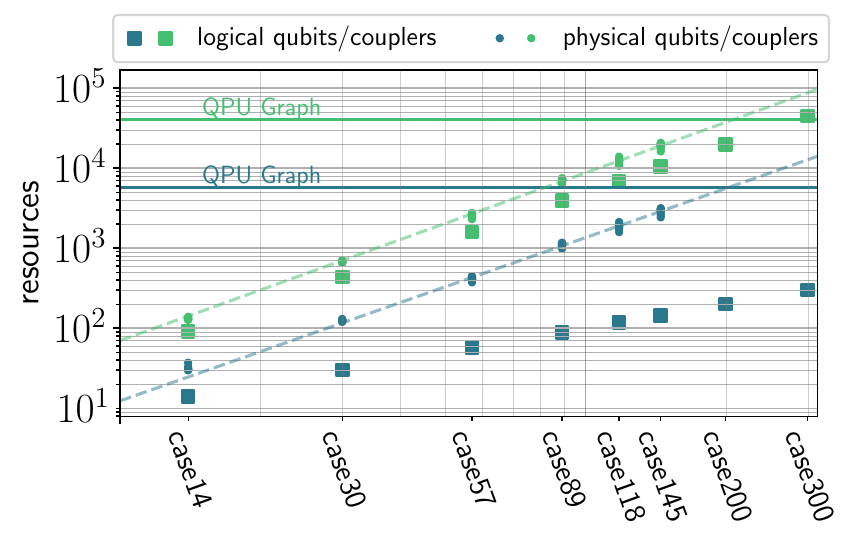}
    \caption{
    Scaling of the number of logical (rectangles) and physical (circles) qubits (blue) and couplers (green) for the \ac{QUBO} $Q_{part.}$ against the grid size of several IEEE test cases given by the number of buses. The physical resources can be estimated by power laws (dashed lines). The hardware limits are drawn as horizontal lines. 
    }
    \label{fig:par:scaling_QUBO}
\end{figure}

By construction, the number of logical variables $N_{q,log}$ is given by the number of buses $N$ in the grid. $Q_{comp.}$ \eqref{eq:par:QUBO_sum_costs} and $Q_{net.}$ \eqref{eq:par:QUBO_network_costs} are fully connected \ac{QUBO} objectives, i.e., each logical qubit has a nonvanishing coupling to every other logical qubit. Thus the full \ac{QUBO} has $N_{c,log} = N(N-1)/2$ logical interactions between logical qubits.  

Since complete graphs, and thus fully connected \acp{QUBO}, can only be directly embedded for $N \leq 4$ on the D-Wave Advantage System 5.4 (JUPSI) with Pegasus topology \cite{dwave2019Advantage, dwave2022Advantage}, the actual resources exceed the logical ones as we have to find minor embeddings for all test cases. We heuristically sample 10 minor embeddings for each test case to generate some minimal statistics. For the 200 and 300 bus test cases no embedding could be found in a reasonable time. For the smaller test cases, we find that the number of physical qubits $N_{q,phy.}$ approximately follows a power law $N_{q,phy.} \approx 0.13 N^{2.01}$ in the number of logical qubits $N$. The power law scales similarly to the number of logical interactions $N_{c,log}$. Likewise, the number of physical couplers $N_{c,phy.}$ needed to embed the problem follows a similar power law $N_{c,phy.} \approx 0.63 N_{c,log}^{2.07}$ in the number of logical interactions, see Fig.~\ref{fig:par:scaling_QUBO}. 

Both power laws estimate resources for other system sizes. The power laws suggest that the current hardware limits are reached at around 200 buses. This is consistent with our findings that no embeddings were found for 200 buses. 

\subsection{Dependence of Solution Quality on Annealing Hyperparameters}
\label{sec:par:hyper}

\begin{figure*}
\centering
\includegraphics[width=\textwidth]{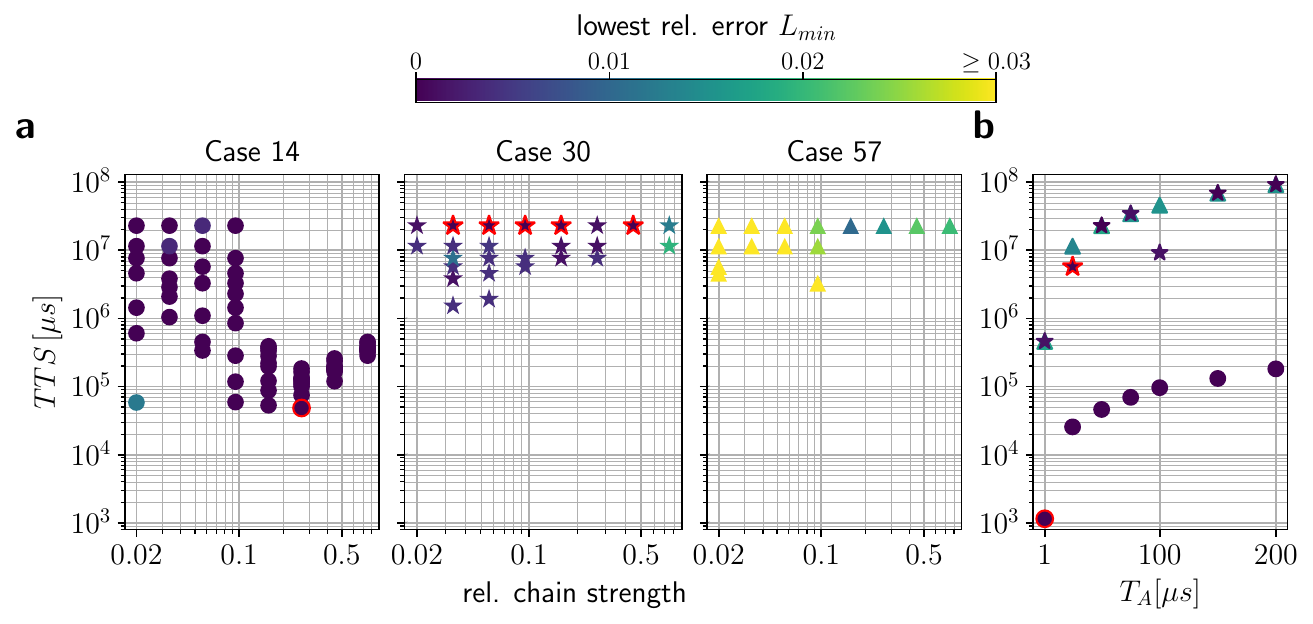}
\caption{Dependence of solution quality on quantum annealing hyperparameters. 
\textbf{a}:  The three panels show the dependence of the $TTS$ for $p_S=0.99$ and $L_{min}$ on the $RCS$ and the different embeddings for cases 14, 30 and 57. If multiple embeddings perform equal in terms of the $TTS$, only the best according to $L_{min}$ is shown here. The full data is shown Fig.~\ref{fig:par:3dplot_chain_strength} in the Appendix. The best hyperparameter configurations with $L_{min} = 0$ are highlighted in red.
\textbf{b}: Dependence of the $TTS$ together with $L$ on the annealing time $T_A$ for the optimal embedding and chain strength combination found in \textbf{a}.}
\label{fig:par:solution_quality}
\end{figure*}

The quality of the sampled solution on the QPU depends strongly on the annealing hyperparameters. We evaluate the quality of a sample set by considering the time to the best solution $TTS$\eqref{eq:par:TTS} for $p_S=0.99$ together with the lowest relative error $L_{min}$ \eqref{eq:par:lowest_rel_error} of the sample set. Here, the energy of the best solution found by Gurobi is used as the reference solution. Hence, if $L_{min}=0$, the best solution found by \ac{QA} is equivalent to the optimal solution found by Gurobi. No optimal solution is obtained if $L_{min}>0$. To generate robust statistics, the sample set for each hyperparameter configuration consists of $N_s = 10^5$ samples.

Similar to the observations made by Gonzalez Calaza et al. \cite{gonzalez_calaza_garden_2021}, we find that the solution quality is sensitive to the used embedding and the \emph{chain strength} $CS$, that is, the coupling strength of the physical qubits in a chain, see \reffig{fig:par:solution_quality}\textbf{a}. For this analysis, we set the annealing time to $T_A = 50 \mu s$ and use the same 10 minor embeddings as in section \ref{sec:par:scaling}. Note that if two embeddings have the same $TTS$ for the same chain strength, only the lower $L_{min}$ is shown \reffig{fig:par:solution_quality}\textbf{a}. 
The full data, showing the performance across all embeddings, is displayed in  Fig.~\ref{fig:par:3dplot_chain_strength} in the Appendix. Furthermore, for better comparison between the test cases, we consider the \emph{relative chain strength} \cite{gonzalez_calaza_garden_2021} given by 
\begin{equation*}
    RCS = \frac{CS}{\max_{(n,m)} \lvert (Q_{part.})_{n,m} \rvert}. 
\end{equation*}
For case 14, the fastest $TTS$ over all ten embeddings is given for $RCS = 0.27$; see the left panel in \reffig{fig:par:solution_quality}\textbf{a}. The performance of the embeddings varies significantly, especially for small $RCS <0.1$. The optimal solution could not be found for some embeddings and small $RCS$, causing the $TSS$ to reflect only the time needed to find the best solution. For all other configurations, the $TTS$ roughly follows a convex function in the $RCS$. The convex shape indicates that if the chain strength is too small, the chains break frequently, preventing them from uniquely representing a logical bit. On the other hand, if the chain strength is too high, the dynamics during annealing are dominated by the internal dynamics of the chains. 

Solving $Q_{part}$ for case 30 to optimality with \ac{QA} is currently at the limit of what can be achieved within a competitive time frame. c.f. the middle panel in \reffig{fig:par:solution_quality}\textbf{a}. For the 8 out of 80 hyperparameter configuration instances where the optimal solution was found, the $TSS$ is $23.03 s$, that is, the optimal solution was sampled only once. On the other hand, for most hyperparameter configurations, no optimal solution has been found ($L_{min} > 0$), although $L_{min}$ remains relatively small. However, regardless of the $RCS$ and across all embeddings, the best (near-optimal) solutions have only been sampled a few times, manifesting in the uniformly slow $TSS$ values. Inspecting the full data, c.f. Fig.~\ref{fig:par:3dplot_chain_strength} in the Appendix, reveals that the optimal solution could only be found for five out of ten embeddings, highlighting the impact of different embeddings on performance. The best performance is achieved for $RCS = 0.15$; optimal solutions have been sampled for four different embeddings. However, the lowest average $L_{min}$ excluding the optimal solutions is achieved for $RCS = 0.95$. 

For case 57, the current hardware limits are exceeded, at least as a global optimal solver. No optimal solutions could be found for any embedding and chain strength. The lowest average $L_{min}$ is achieved for $RCS = 0.15$.  Additionally (not shown in the figure), we sampled the QPU $10^6$ times for the best-performing combination of $RCS$ and embedding for case 57, which did not significantly improve the solution quality. 

Another hyperparameter that, by definition, affects the $TTS$ is the annealing time $T_A$. For the best embedding and chain strength combinations from the previous analysis, the $TTS$ increases in $T_A$ for all three cases; see \reffig{fig:par:solution_quality}\textbf{b}. For case 14, the optimal solution is found for all values of $T_A$. Hence, the smallest problems, which do not seem to need extensive hyperparameter optimization, could profit from using the fast annealing feature, which just became available, to achieve an even smaller TTS. On the other hand, for case 30 optimal solutions are only sampled for $T_A \geq 25 \mu s$. The optimal performance as a global solver is achieved for $T_A = 25 \mu s$. Although for case 57 the optimal solution is never found, the lowest $L_{min}$ is achieved for $T_A = 50 \mu s$, already while scanning for chain strengths, see \reffig{fig:par:solution_quality}\textbf{a}. The adiabatic theorem supports the observation that for harder problems, that is problems with a smaller minimum energy gap, longer $T_A$ leads to better results since the probability of jumping to a higher energy state throughout the annealing process is reduced. 

Other hyperparameters, in particular anneal offsets, have also been found to be important for graph partitioning \cite{barbosa_optimizing_2021}. It has been demonstrated that tuning anneal offsets on a qubit level can prevent chains from freezing out early during the anneal \cite{andriyash2016} and to prevent perturbative anticrossings \cite{lanting_experimental_2017}. However, since tuning anneal offsets is quite resource-intensive, we leave this for future work.  

\begin{table}[t!]
\centering
\begin{tabular}{||c || c | c || c | c||} 
    \cline{2-5}
      \multicolumn{1}{c|}{} & \multicolumn{2}{c||}{Gurobi} & \multicolumn{2}{c||}{\ac{QA}} \\
    \multicolumn{1}{c|}{} & \multicolumn{2}{c||}{} & \multicolumn{2}{c||}{(best hyperparamter)} \\ [0.5ex] 
 \cline{2-5}
    \multicolumn{1}{c|}{} & MIP gap & $T_{\text{Gur.}}$ [$s$] & $TTS$[$s$] & $L_{min}$\\ 
 \cline{2-5} 
 \hline
    Case 14 & 0 & 0.43 & 0.05 & 0 \\ 
    Case 30 & 0 & 4.66 & 23.03 & 0 \\ 
    Case 57 & 0 & 1.04 & 23.03 & 0.01 \\ 
 \hline  \noalign{\vskip 0.03in} 
\end{tabular}

\caption{Comparison of results obtained by Gurobi and \ac{QA} ignoring all overheads for \ac{QA}.}
\label{tab:par:solver_comp}
\end{table}

We conclude that the performance of \ac{QA} depends strongly on hyperparameters, particularly on the embedding and the chain strength. Furthermore, for quantum annealing, the actual time to solution suffers from overheads, i.\,e. the time needed to embed the problem and network latency. For a fair comparison between the performances of quantum annealing on current prototypical hardware with classical deterministic or heuristic solvers, the overhead must be included, and the question arises whether to compare optimized solvers. A fair comparison would thus reveal current hardware limitations of the D-Wave machine, in particular, the need for minor embeddings \cite{vert2024benchmarking}, and is thus not meaningful for the future where some problems might have been overcome, e.g., by adapting the driver to expand the minimal gap \cite{hen_quantum_2016}. 
Nevertheless, ignoring all aforementioned overheads and comparing the results obtained by \ac{QA} for the best hyperparameters with Gurobi, we observe that \ac{QA} ``outperforms" Gurobi running on a laptop only for case 14, see Table~\ref{tab:par:solver_comp}. This observation is consistent with the recent literature: For other highly connected Ising problems, quantum annealing on current D-Wave machines does not provide a speed up or better solution quality \cite{vert2024benchmarking, au2023np, mohseni2022ising}. 

\subsection{Near Optimal Solutions}

\begin{figure}
\centering
\includegraphics[width=\columnwidth]{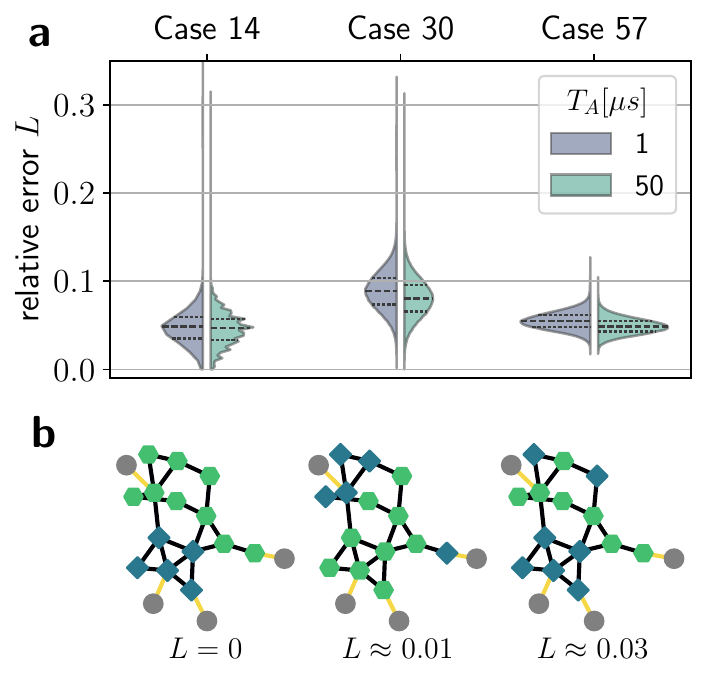}
\caption{Assessment of near-optimal solutions. 
\textbf{a}: Comparison of the distributions of $L$ between different test cases and annealing times $T_A$. \textbf{b}: Near-optimal partitionings ($L>0$) for the 14 bus test case in comparison to the optimal partitioning ($L = 0$).}
\label{fig:par:near_optimal}
\end{figure}

For larger grids and using non-optimal hyperparameters, we are likely to only sample near-optimal solutions $L>0$ on the QPU. For distributed parallel simulation, all partitionings are, in principle, feasible solutions. If the partitioning is done in the loop, the total runtime is the sum of the overhead due to optimizing the partitioning, i.\,e. determined by the number of samples $N_s$, the annealing time $T_A$, and the runtime of the distributed simulation. Instead of solving the grid partitioning \ac{QUBO}~\eqref{eq:par:Qubo_grid_part} up to optimality, which for large grids becomes very hard, sampling near-optimal solutions very fast might be a viable near-term application of quantum annealing. The trade-off between shorter overheads due to sampling fewer solution candidates and longer simulation times due to sub-optimal partitionings needs to be investigated in a separate study. The optimal number of samples will also depend on how often we need to (re-)partition the network. 

Furthermore, the optimal number of samples also depends on the annealing time $T_A$. As a preliminary step, we investigate the distribution of the relative error $L$ for the best embedding and chain strength combinations found in the previous section for different $T_A$, cf.~Fig.~\ref{fig:par:near_optimal}\textbf{a}. We find that the mean and the quantiles of the distributions of $L$ are, as expected, lower for $T_A = 50 \mu s$ than for $T_A = 1 \mu s$. Hence, the optimal $T_A$ for the whole loop can not be assessed by just looking at the $TTS$ for the optimization problem alone. Interestingly, the mean and variance of the distributions of $L$ for case 57 are lower than those for case 30, although solving case 57 up to optimality is harder than solving case 30 for \ac{QA}. We assume that this alleged paradox can be resolved by comparing the topologies and generator locations of case 30 and case 57, as shown in Fig.~\ref{fig:par:case30_optimal_solutions}. In case 57, the generators are all concentrated in the ``southwest" of the grid. As a result, any partition that seeks to balance both the components, particularly the costly generators and the sizes of the sub-networks, must cut diagonally from the southwest to the northeast. Furthermore, the northeast region consists of only a few large cycles, further limiting ``good" partition choices. 
This translates to a significant gap between the energy (objective value) levels of $Q_{part.}$ corresponding to ``good" cuts along the southwest-to-northeast axis and the energy levels for all other ``bad" cuts, where the generators would be strongly imbalanced. Therefore, during the anneal, the system is unlikely to get excited to the higher energy levels associated with ``bad" cuts along any other axis. On the other hand, case 30 is more homogeneous, and thus we expect that energy levels are more evenly distributed. Consequently, transitions or cascades of transitions to higher energy levels corresponding to ``bad" partitionings are more likely. 

In general, we can hope that sampling partitionings on current (or near-term) QPUs for larger grids will also return near-optimal solutions that are good enough to provide simulation speed-ups in competitive time. However, near-optimal solutions might be less intuitive than optimally partitioned sub-networks. For case 14, first, the cut becomes larger with increasing $L$; however, the components and sub-network sizes remain (roughly) balanced, cf. Fig.~\ref{fig:par:near_optimal}\textbf{b}. The increase in cut size leads to disconnected sub-networks. 

\subsection{Rescaling the QUBO to improve performance}
\label{sec:par:rescaling_qubo}

The three QUBO objective functions of $Q_{part}$ have been designed to estimate the overhead or (squared) idle time associated with each of the three steps in every iteration of the parallel simulation. So far, $Q_{part}$ is a simple sum of the three objectives, effectively balancing the overheads and squared idle times, see Eq.~\ref{eq:par:Qubo_grid_part}. However, in the context of linear scalarization (cf. Eq.~\ref{eq:par:single_objective}), the individual objectives can be rescaled by introducing scalars as 
\begin{equation*}
    Q_{part}(\vec{\lambda}) = \lambda_{comp. }Q_{comp.} +  \lambda_{cut}Q_{cut} +  \lambda_{net.} Q_{net.}.
\end{equation*}
Rescaling changes the importance of the single objectives and altering the energy landscape of the problem and thus eventually the optimal partitioning. On the other hand, any reformulation, and thus any rescaling of the single objectives, can also influence the solution time of solvers involving heuristics. For \ac{QA} rescaling can modify the separation of the energy levels and thus transition probabilities to excited states. 

\begin{figure}
\centering
\includegraphics[width=\columnwidth]{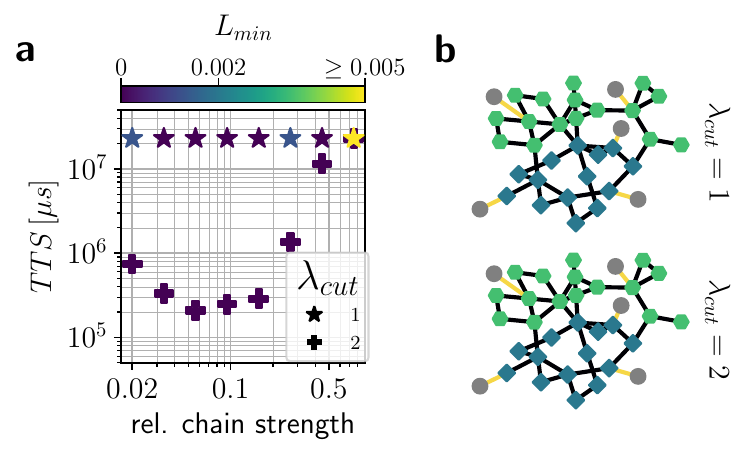}
\caption{Effect of rescaling the energy of $Q_{cut}$ for case 30. 
\textbf{a}: Comparison of the performance ($TTS$) and solution quality ($L_{min}$) obtained by \ac{QA} between $\lambda_{cut}=1$ and $\lambda_{cut}=2$. For each $RCS$ value only the minimal $L_{min}$ with the fastest $TTS$ among all 10 embeddings is shown here. \textbf{b}: Optimal partitioning for $\lambda_{cut}= 1$ and $\lambda_{cut}= 2$.}
\label{fig:par:rescaling_qubo}
\end{figure}

For case 30, we observe that increasing the importance of $Q_{cut}$, in particular setting $\lambda_{cut} = 2$ while keeping $\lambda_{comp. \backslash net.} = 1$, significantly improves the performance of \ac{QA}, see Fig.~\ref{fig:par:rescaling_qubo} \textbf{a}, while the optimal partitioning is not affected, see subfigure \textbf{b}. For $\lambda_{cut} = 2$ the optimal solution is found for all tested chain strengths. The fastest $TTS$ is $0.21s$, which constitutes an improvement by a factor of 110 compared to the fastest $TTS$ with $\lambda_{cut} = 1$. Further research is needed to test the effect of rescaling across different test cases, as well as, the trade-off between performance and optimal partitionings. 

\section{Conclusion and Outlook}
\IEEEpubidadjcol

This article presents a novel quantum optimization approach to optimal grid partitioning for diakops-based parallel simulation approaches for power grids. 
We derived a \ac{QUBO} formulation by mapping the problem-specific objectives, i.\,e. minimizing the idle times and the overhead due to the cut, to general \ac{QUBO} objectives for graph partitioning; an approach that might be repeated for other grid partitioning applications \cite{sanchez2013partitioning, sanchez_gracia2014spectralclustering, weiqing2008statediakoptics}.
The \ac{QUBO} scales linearly in the number of buses and is thus efficient for optimal hardware. Moreover, the optimal solutions to the \ac{QUBO} formulation provide reasonable partitionings for some small test cases. 

Furthermore, for current quantum hardware, we have demonstrated that 
\begin{enumerate}
    \item the partitioning of  only small grids can be optimized since the \ac{QUBO}s connectivity requires an embedding on the QPU graph, 
    \item the $TTS$ and the quality of the solutions depend on the grid size such that for a test case with 57 buses no optimal solution can be found in reasonable time, and
    \item the $TTS$ and the quality of the solutions are highly sensitive to annealing hyperparameters such as the embedding and the annealing time. 
\end{enumerate}
These problems are on such a fundamental level, that benchmarking the quantum solver against classical deterministic optimization approaches is, in our view, not meaningful. The current quantum annealers are too noisy and too sensitive to the choice of hyperparameters. 

In our future research, we plan to investigate whether sampling near-optimal solution fast might be preferred over opting for optimal solutions to the \ac{QUBO} when accounting for the whole partitioning-simulation loop. This is of interest, especially if network topologies change frequently. Whether \ac{QA} might then offer an advantage over other heuristic approaches, c.f.\cite{kyesswa_2020_partioning}, must be studied.  

We are also planning to investigate whether it is sensible to formulate a more complicated objective function, e.g. to enforce connectivity of the sub-networks or to include modularity to limit solver iteration, hoping to achieve more optimal overall performance considering the partitioning together with the parallel simulation. 

\section{Code Availability}

The data and code that support the findings of this article are openly available~\cite{code_publication}.

\section{Acknowledgements}
The paper was written as part of the project ``Quantum-based Energy Grids (QuGrids)", which is receiving funding from the programme ``Profilbildung 2022", an initiative of the Ministry of Culture and Science of the State of North Rhine-Westphalia. The authors gratefully acknowledge the Jülich Supercomputing Centre (\url{https://www.fz-juelich.de/en/ias/jsc/systems/supercomputers/apply-for-computing-time/juniq}) for funding this project by providing computing time on the D-Wave Advantage™ System JUPSI through the Jülich UNified Infrastructure for Quantum computing (JUNIQ), which has received funding from the German Federal Ministry of Education and Research (BMBF) and the Ministry of Culture and Science of the State of North Rhine-Westphalia.
The authors also thank Manuel Dahmen for the fruitful discussions. 
The sole responsibility for the content of this publication lies with the authors.

\section{Appendix}

\begin{figure}[H]
    \centering
    \includegraphics[width=\columnwidth]{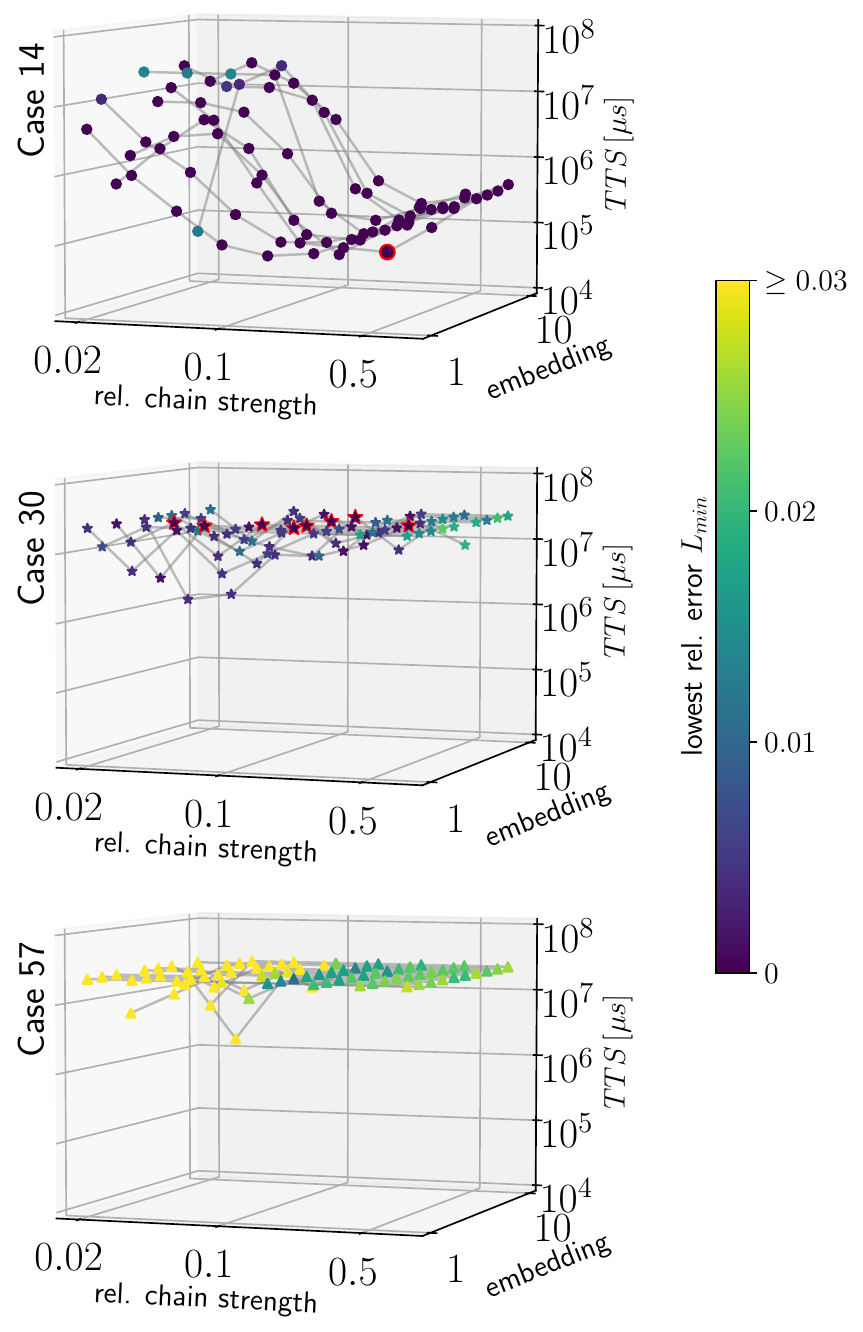}
    \caption{Solution quality across different embedding and chain strength combinations for cases 14, 30, and 57. The data has been ``projected" into a 2d plane for Fig.~\ref{fig:par:solution_quality}~\textbf{a}.  }
    \label{fig:par:3dplot_chain_strength}
\end{figure}

\bibliographystyle{IEEEtran}

\bibliography{references_submission_review}

\end{document}